\documentclass[
prb,twocolumn,10pt,
superscriptaddress,
 amssymb,amsmath,
 aps,
 footinbib, amsart
]{revtex4-1}

\usepackage{pagenote}

\makepagenote
\usepackage{multirow}
\usepackage{tcolorbox}
\usepackage{graphicx}
\usepackage{overpic}
\usepackage{dcolumn}
\usepackage{cancel} 
\usepackage{amssymb}
\usepackage[normalem]{ulem}
\usepackage{bm}
\usepackage{subfigure}

\begin{document}

\newcommand{\td}[1]{{\color{orange}{#1}}}
\newcommand{\AG}[1]{{\color{violet}{#1}}}
\newcommand{\LT}[1]{{\color{cyan}{#1}}}
\newcommand{\change}[1]{{\color{black}{#1}}}
\newcommand{\typo}[1]{{\color{red}{#1}}}

\newcommand{\bto}{BaTiO$_3$}
\newcommand{\kno}{KNbO$_3$}
\newcommand{\pto}{PbTiO$_3$}

\newcommand{\pa}{paraelectric}
\newcommand{\ma}{M$_\text{A}$}
\newcommand{\mb}{M$_\text{B}$}
\newcommand{\mc}{M$_\text{C}$}

\newcommand{\oop}{out-of-plane}
\newcommand{\ip}{in-plane}

\newcommand{\pol}[1]{$P_{[{#1}]}$} 

\newcommand{\sbs}{side-by-side}
\newcommand{\lbl}{layer-by-layer}
\newcommand{\ep}{experiment}
\newcommand{\sg}{single domain}
\newcommand{\unitp}{$\mu$C/cm$^2$}

\newcommand{\uef}[1]{~kV/cm}
\newcommand{\dir}[1]{[{#1}]}   
\newcommand{\dirs}[1]{$\langle${#1}$\rangle$}   
\newcommand{\pla}[1]{({#1})}   
\newcommand{\plas}[1]{\{{#1}\}}   

\title{Competing phases and domain structures of ferroelectric perovskites: the benefit of epitaxial (110) growth}

\author{Lan-Tien Hsu}
\email{lan-tien.hsu@ruhr-uni-bochum.de}
\affiliation{
 Interdisciplinary Centre for Advanced Materials Simulation (ICAMS), Center for Interface-Dominated High Performance Materials (ZGH)\change{,} and Faculty for Physics and Astronomy, Ruhr-University Bochum, Universitätsstr. 150, 44801 Bochum, Germany
}
\author{Takeshi Nishimatsu}
\author{Anna Grünebohm}
\email{anna.gruenebohm@ruhr-uni-bochum.de}
\altaffiliation[Also at ]{ICAMS, ZGH, and Faculty for Physics and Astronomy, Ruhr-University Bochum}

\begin{abstract}

Strain and domain engineering offer powerful routes to control phase and domain stability in ferroelectric thin films. While most studies have focused on (100)-oriented growth, the impact of lower-symmetry orientations remains underexplored. We address this knowledge gap with first-principles-based molecular dynamics simulations for the example of prototypical ferroelectric perovskites under (110) strain. Epitaxial (110) strains may indeed outperform the widely studied (100) orientation, as even modest strain values stabilize a diverse set of metastable nanoscale states with potential high functional tunability. In this regime, the films exhibit multidomain configurations with domain wall normals oriented along the clamped in-plane or the relaxed out-of-plane directions and heterophases in BaTiO$_3$ and KNbO$_3$. Besides, complex superdomain patterns and antiferroelectric-like domains are observed in PbTiO$_3$. These metastable nanoscale configurations may allow for large reversible responses.\\
\textbf{keywords: ferroelectrics, domains, strain engineering, phase transitions, molecular dynamics simulations, oxide perovskites}

\end{abstract}

\maketitle

\section{Introduction}

Ferroelectrics are utilized for applications ranging from tunable capacitors and high-frequency applications to non-volatile memory devices, and are promising candidates for solid-state cooling and neuromorphic computing.\cite{grunebohm_interplay_2021, iniguez-gonzalez_creating_2024,park_overcoming_2023,martin_thin-film_2016} These applications require large functional, i.e.,\ dielectric, piezoelectric, or electrocaloric responses,\cite{whatmore_100_2021}  either related to electric field-induced phase transitions or the properties and mobility of dense domain walls.\cite{grunebohm_interplay_2021} Thus, the understanding and precise control on phases and domain structures is essential.

In this quest, strain-engineering in epitaxial films\cite{dhole_strain_2022,li_insights_2024,schlom_strain_2007} may change \change{their} transition temperatures by hundreds~K, and can stabilize either new \sg\ or multidomain states, depending on the anisotropy  of \change{ elastic constants of the material}.\cite{damjanovic_contributions_2005,xue_strain_2016,bratkovsky_effects_2011}
Particularly, tensile and compressive strain stablizes \ip\ and \oop\  polarization, respectively, and may act as restoring force for reversible polarization rotation. \change{These restoring forces may} consequently adjust the remanent polarization and coercive field.\cite{pal_subsecond_2025, van_truong_enhancement_2023,zhang_tuning_2009,choi_enhancement_2004}
Besides, \change{materials} can also accommodate strain  by the  coexistence of different phases and domains \change{in} so-called ``heterophases''.\cite{topolov_heterogeneous_2018,roytburd_stability_2011} \change{This mechanism is so far not widely discussed in the context of strained ferroelectric perovskite oxide films.} 
All these low symmetry phases\cite{gui_properties_2011,vanderbilt_monoclinic_2001}, multidomain structures\cite{qiao_dielectric_2008, chatterjee_interplay_2024,ren_bimodal_2022,tovaglieri_investigating_2025,matzen_super_2014,li_thickness-dependent_2017,yasui_complex_2012}, and phase mixtures\cite{roytburd_stability_2011}  may allow \change{for} large or tunable reversible responses, e.g.,\  dielectric, piezoelectric, and electrocaloric responses, as well as negative capacitance\cite{salahuddin_use_2008,zubko_negative_2016,marathe_electrocaloric_2014,imai_orientation_2010,chen_colossal_2025}.

So far, the main focus on research has been  the high symmetric  growth direction $\langle 001 \rangle$. Already for that, it has been reported that strain can bring large changes in phase diagram. \change{For example,}  the transition temperature between the paraelectric and ferroelectric states can increase \change{by} about 30~K under only 0.1\% epitaxial (001) strain.\cite{grunebohm_ab_2015} 
Even larger modifications are expected when symmetry is further reduced, e.g.,\ using (110) as growth directions, a regime that remains surprisingly underexplored. In the few existing  studies,  triclinic and monoclinic low-symmetry phases, and  multidomain structures have been reported.\cite{wu_effect_2016, ma_effect_2017,raeder_anisotropic_2021,ren_bimodal_2022}

So far,  theoretical studies on the misfit-temperature  phase diagram, the so called ``Pertsev" diagram\cite{pertsev_effect_1998}, of epitaxially \pla{110} strained ferroelectric perovskites are underrepresented in literature: phenomenological models\cite{wu_effect_2016, ma_effect_2017} based on 
a priori assumption of single domain, phase-field models\cite{li_misfit_2024,zhang_strain_2022}, which are not limited by such assumptions but are parameter-sensitive\cite{wang_variation_2018}, and only one  study based on an effective  Hamiltonian\cite{gui_properties_2011} have been reported.
\change{The latter study has however} focused on \sg\ structures, and a systematic comparison between ferroelectric perovskite oxides is still missing. 
Therefore, we revisit the influence of epitaxial \pla{110} strain on phase and domain structure of ferroelectric perovskites using  a first-principles based effective Hamiltonian. 
We  compare the two protoypical materials \bto\ and \kno\ which have similar phonon spectrum of their \pa\  cubic (C, space group: $Pm\bar3m$) phase\cite{ghosez_lattice_1999}, and both undergo transitions from \pa\ cubic  to ferroelectric tetragonal (T, $P4mm$) to orthorhombic (O, $Amm2$) to rhombohedral (R, $R3m$) phase when cooling without strain. \change{The transition temperatures however differ} (\bto\cite{aksel_advances_2010}: 393~K, 278~K, and 183~K; \kno\cite{you_synthesis_2021}: 702~K, 489~K, and 260~K).
We contrast these findings to  \pto, which undergoes one  spontaneous transition at 753~K\cite{yoshiasa_high-temperature_2016}  from \pa\ C to ferroelectric T phase.

We reveal the potential of (110)-oriented growth\change{, which} can induce a multitude of uncommon domain structures, including (meta)stable domain configuration for $\eta \sim 0$,  heterophases in a large temperature range, and domain walls parallel to the interface for \bto\ and \kno\ \change{as well as} antiferroelectric-like domains and nanosized superdomains in \pto.

\section{Method}
We performed coarse-grained molecular dynamics simulations  in the canonical ensemble with the Nos\'e-Poincar\'e thermostat\cite{Bond.1999} using a simulation box of 48$\times$48$\times$48 unit cells (side length: around 20~nm), if not otherwise stated. During the simulations, periodic boundary conditions were applied in all directions, and a time step of 2~fs was used. The simulation and post-processing were done with the  open-source packages {\it{feram}}\cite{Nishimatsu.2008} and  {\it{AutoFeram}}\cite{icams-sfc_2024}. 

The formulation of the used  effective Hamiltonian by Zhong {\it{et al.}}\cite{zhong_phase_1994,zhong_first-principles_1995}, parameterized by first-principles density functional theory calculations for \bto\cite{nishimatsu_first-principles_2010}, \kno\footnote{T.\ Nishimatsu, feram at sourceforge.net (2015)}, and \pto\cite{nishimatsu_molecular_2012,waghmare_ab_1997}, is given as: 
\begin{equation} \label{eq:Eeff}
\begin{aligned}
H^\text{eff} &= E^\text{k,dipole}(\{\bm{u}\}) + 
V^\text{self}(\{\bm{u}\})  
+ V^\text{dpl}(\{\bm{u}\}) 
\\& + V^\text{short}(\{\bm{u}\}) 
+ V^\text{elas,homo}(\{{\eta_1,...,\eta_6}\})
\\& + V^\text{elas,inho}(\{\bm{w}\}) 
+ V^\text{coup,homo}(\{\bm{u}\},\{{\eta_1,...,\eta_6}\})
\\& + V^\text{coup,inho}(\{\bm{u}\},\{\bm{w}\}) 
- Z^*_\text{eff} \sum_{R} \bm{E  \cdot u(R)}\,,
\end{aligned}
\end{equation}
where  $\{\bm{u}\}$ and $\{\bm{w}\}$  are the local soft-mode and acoustic displacement vectors of the unit cells, $\{\bm{\eta_l}\}$ is the homogeneous strain tensor of the whole system in Voigt notation,  $Z^*_\text{eff}$ is the effective charge associated with the soft mode, and  the local polarization vector $\bm{p}(\bm{R})$ of the unit cell at  position $\bm{R}$ is given as 
$\bm{p}=Z^*_\text{eff}$~$\bm{u}(\bm{R})/\Omega$, with $\Omega$ being the unit cell volume. The potential energy includes: the on-site local mode self energy ($V^\text{self}$), the long-range dipole-dipole Coulomb interaction ($V^\text{dpl}$), the short-range local soft mode interaction ($V^\text{short}$), the elastic energy from homogeneous ($V^\text{elas,homo}$) and inhomogeneous strains ($V^\text{elas,inho}$), the coupling between the local soft mode and the homogeneous strain ($V^\text{coup,homo}$) and between the local soft mode and the inhomogeneous strain ($V^\text{coup,inho}$), and the electric enthalpy ($Z_\text{eff} \sum_{R} \bm{E \cdot u(R)}$), which reflects the \change{coupling between electric field and polarization}, see Ref.~\onlinecite{Nishimatsu.2008} for further details. The strain is internally optimized and  only the kinetic energy of the local soft modes ($E^\text{k,dipole}$),  is 
explicitly treated in MD simulations.
While this formulation underestimates  $T_c$ \change{partly} due to  neglected anharmonic couplings to higher-energy phonon modes,\cite{Nishimatsu.2010, mayer_improved_2022} it can correctly predict changes of phase stability with strain or composition, domain structures, and   functional responses.\cite{mayer_hidden_2023, grunebohm_influence_2023,lisenkov_tuning_2018, gui_tuning_2015,walizer_finite-temperature_2006,kumar_domain_2013,kumar_first-principles_2010}

\change{Note that  potential instabilities related to antiferrodistortive octahedral rotations are neglected. While the phonon spectra of the paraelectric phase of BaTiO$_3$ and KNbO$_3$ do not have such instabilities with or without strain, see Fig.~\ref{fig:phonon spectrum}~(a) and (b), a competing antiferrodistortive instability is present in PbTiO$_3$\cite{wojdel_first-principles_2013}. However, this instability is weaker compared to the polar phonon mode and not sensitive to strain. 
 Although  an additional instability for in-phase octahedral rotation ($\vec{q}=[1,0,1]$ and $[0,1,1]$) occurs under compressive (110)-strain, it is  not dominant in the strain range of interest, see blue and red crosses in Fig.~\ref{fig:phonon spectrum}~(c).
 The main trend of transition temperatures and local and global polarization are covered by the model for all three materials, while potential additional octahedral rotations  in PbTiO$_3$ at low temperatures are  neglected.\cite{wojdel_first-principles_2013,nishimatsu_first-principles_2010}
}
\change{Further n}ote that the chosen parametrization for \pto\ \cite{waghmare_ab_1997,nishimatsu_molecular_2012}  overestimates the elastic constant C$_{44}$ \change{from literature} by a factor of four \cite{liu_first-principles_2008,king-smith_first-principles_1994,piskunov_bulk_2004}. However, the observed domain configurations and the wall  periodicity are not modified when the parametrization is modified correspondingly. 
\footnote{The (110)-induced domain structures of PbTiO$_3$ turn out to not be sensitive to C$_{44}$, which represents the resistance to shear deformation. Even if we reduce this elastic constant  by a factor of four and adjust all directly depending parameters, no qualitative changes are observed.}

\begin{figure}
    \centering
    \begin{overpic}[width=0.48\textwidth, trim=0cm 0.5cm 2cm 9cm,clip]{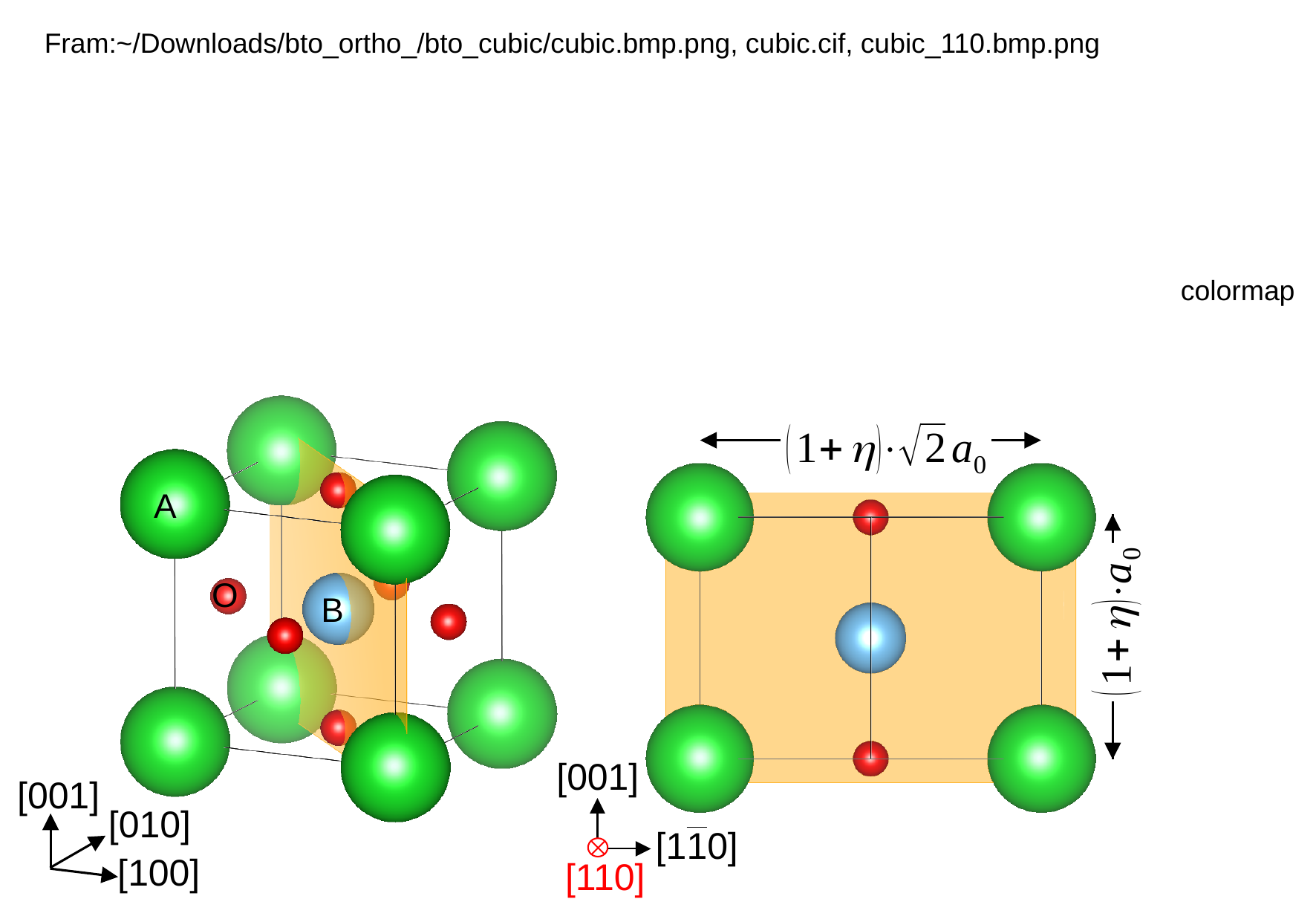}
    \put(0, 36){\text{\large(a)}}
    \put(48,36){\text{\large(b)}}
    \end{overpic} 
    \caption{(a) Unit cell of the ABO$_3$ \typo{perovskite} structure (A (green): Pb, Ba, K; B (blue): Ti, Nb; O (red)) with alternating O$_2^{4-}$ and A-B-O$^{4+}$ (orange)
     \pla{110} planes. (b) The elastic boundary conditions illustrated on a \pla{110} plane.
     The lattice is relaxed along the growth direction, i.e.,\  \dir{110}, and the lattice constants along \dir{001} and \dir{1$\bar1$0} are clamped to $(1+\eta)a_0$ and $(1+\eta)\sqrt{2}a_0$, respectively, where $\eta$ is the external strain and $a_0$ is the reference lattice parameter taken from the \pa\ cubic phase directly above the ferroelectric  transition temperature.}
    \label{fig:unitcell}
\end{figure}

We studied defect-free, perfectly clamped \pla{110}-oriented thin films coherently grown on (110)-oriented cubic substrates  under short-circuit boundary conditions. Therefore, the lattice parameters  were fixed along the  \dir{001} and \dir{1$\bar1$0} directions to $(1+\eta)\cdot a_0$ and $(1+\eta)\cdot \sqrt{2} a_0$, see Fig.~\ref{fig:unitcell}, where $\eta$ is the external strain and $a_0$ is the reference lattice parameter of the \pa\ cubic phase directly above the paraelectric to ferroelectric transition temperature   \change{predicted by the used model (BaTiO$_3$: 3.996~\AA, KNbO$_3$: 4.029~\AA, and PbTiO$_3$: 3.996~\AA).}
and both $\angle$(\dir{001}, \dir{110}) and $\angle$(\dir{001}, \dir{1$\bar1$0}) were fixed to 90$^{\circ}$\change{.}

Neither the impact of alternating charges in BaTiO$^{4+}$ and O$_2^{4-}$ layers, nor the atomistic interface structure and possible depolarization fields were considered.
These boundary conditions have been included in the Github repository of {\it{feram}}.\cite{nishimatsu_feramsrcoptimize-homo-strain-110f_nodate}

Cooling simulations were done on a strain-temperature grid of 0.1~\% and  5~K, starting  well in the \pa\ phase with a homogeneous normal distribution of dipoles (with atomic displacement of mean: 0~\AA\ and  standard deviation: 0.02~\AA).
At each temperature, the system was thermalized for 120~ps and then averaged over the following 40~ps. 
\change{Inspired by Ref.~\onlinecite{rose_giant_2012} we determined the transition temperatures $T_c$ by the maximal changes of the polarization. For that purpose we monitored the change in the distribution of $\bm{p}(\bm{R})$.}

\begin{figure}
    \centering
    \begin{overpic}[width=0.48\textwidth, trim=4cm 9.8cm 10cm 4.8cm, clip]{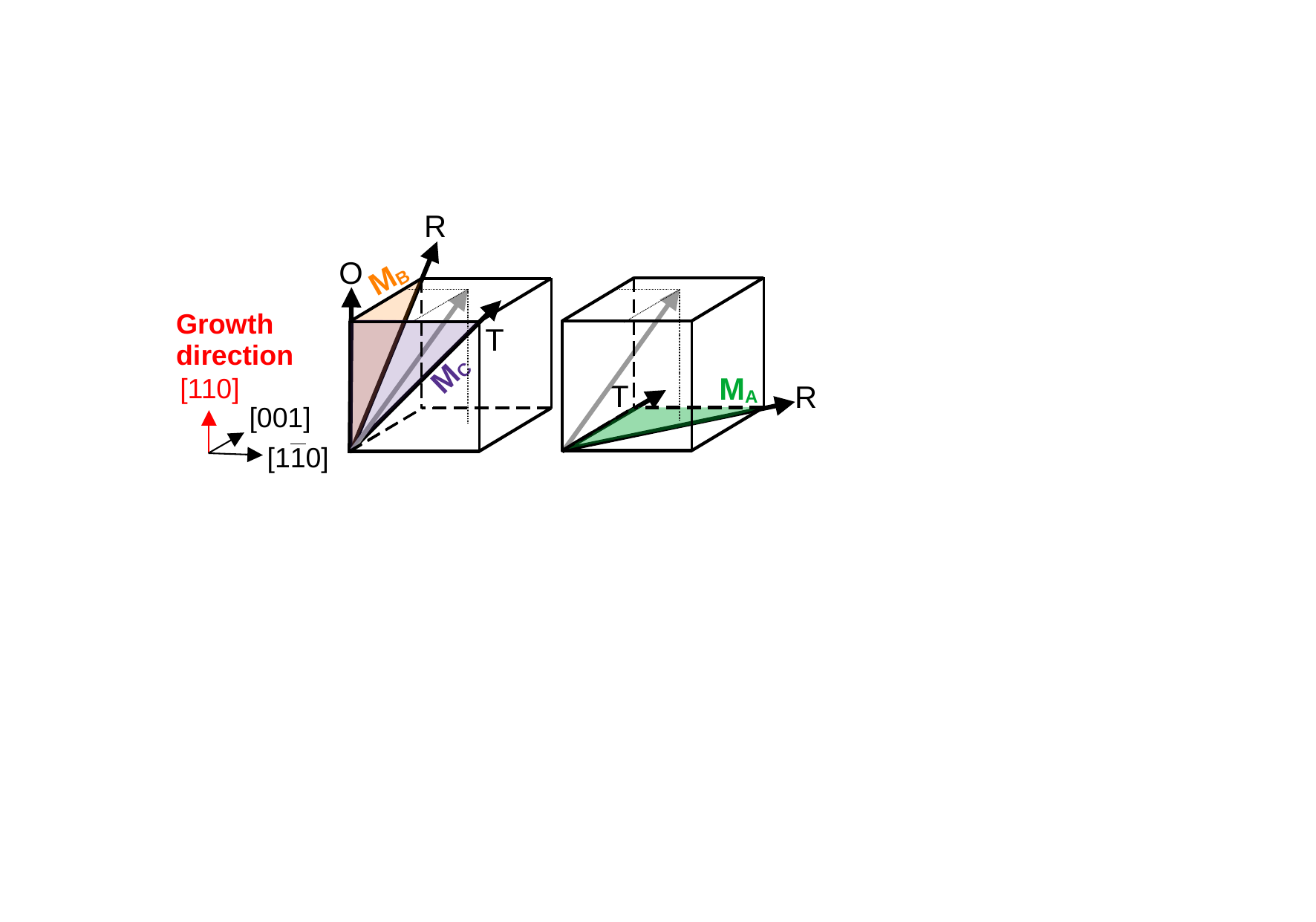}
    \put(18,34){\text{\large(a)}}
    \put(53,34){\text{\large(b)}}
    \end{overpic}   
    \caption{Possible polarization directions relative to the [110] growth direction.  (a) Under compressive strain one polarization component is  along (110), i.e.,\ O phase, and with increasing magnitude of \pol{001} or \pol{1\bar10} \mb\ and \mc\ phases bridge towards R and T phases, respectively.
 (b) Under tensile strain, the in-plane polarization can rotate between T and and R phase via phase \ma. An example of possible Tri phases with finite \pol{001}, \pol{110}, and \pol{1\bar10} are  shown in gray arrows. }
    \label{fig:phasedirection}
\end{figure}

To classify the  (local) polarization, we followed the convention based on the three Cartesian components of $\bm{P}$ ($a > b > c > 0$) by Vanderbilt and Cohen\cite{vanderbilt_monoclinic_2001}: $\langle$a,0,0$\rangle$ for tetragonal phases (T); $\langle$a,a,0$\rangle$ for orthorhombic phases (O); $\langle$a,a,a$\rangle$ for rhombohedral phases (R);  $\langle$b,b,a$\rangle$, $\langle$a,a,b$\rangle$, $\langle$a,b,0$\rangle$ for the monoclinic phases M$_\text{A}$, M$_\text{B}$, and M$_\text{C}$, respectively, and $\langle$a,b,c$\rangle$ for triclinic phases (Tri), see Fig.~\ref{fig:phasedirection}. 
Note that, due to the straining, these directions of $\bm{P}$ may not correspond to the symmetry of the lattice. For example, the system at 400~K under $\eta=0.75\%$ is classified as T phase as the polarization is along the [001] direction, however its lattice constants are $a_{[100],\text{ps}} = a_{[010],\text{ps}} < a_{[001]} = 1.0075 a_0$ and its cell angles are $\angle ([100]_\text{ps}, [001]) = \angle ([010]_\text{ps}, [001]) = 90^{\circ} < \angle ([100]_\text{ps}, [010]_\text{ps})$.

The domain structures were visualized with  Ovito\cite{ovito} 
\change{and we determined the mean local polarization  of each domain $\bm{P}$  from the time-averaged local polarization $\bm{p}(\bm{R})$}. For simplicity, the projection of $\bm{P}$ on the direction [ijk] is  given as ``\pol{ijk}''. 

Domain walls were labeled based on their phase and the angle of the polarization rotation follow the convention by 
 \citeauthor{marton_domain_2010}\cite{marton_domain_2010}, e.g.,\  T180 for the wall separating T domains with polarization antiparallel to each other. To simplify the text, we refered to  wall with normal along [xyz] as ``[xyz]-wall''.

For field ramping, an electric field of \change{0.177}~kV/(cm$\cdot$ps) is applied.

\change{To study the dynamical instabilities of these three materials, we performed density functional perturbation theory (DFPT)\cite{gonze_dynamical_1997} using abinit\cite{troeye_recent_2016}. We used PBEsol exchange-correlation functionals and the optimized norm-conserving pseudopotentials from the PseudoDojo server\cite{hamann_optimized_2013} with Ba $5$s$^25$p$^66$s$^2$, K $3$s$^23$p$^64$s$^1$, Nb $4$s$^24$p$^6$$4$d$^4$$5$s$^1$, O $2$s$^2$p$^4$, Pb $5$d$^{10}$$6$s$^2$$6$p$^2$, and Ti $3$s$^23$p$^63$d$^24$s$^2$ electrons as valence states. The plane-wave energy cut-off was set to 55~Hartree and the Brillouin zone was sampled using a $8\times8\times8$ Monkhorst-pack $k$-mesh\cite{monkhorst_special_1976, chadi_special_1973} for a five-atom supercell with lattice vectors along [110], [001], [1$\bar{1}0$] directions. The self-consistent field and relaxation cycles were converged with a tolerance on the residual potential being $1\cdot 10^{-10}$ Hartree and a tolerance on the maximal force being $5\cdot 10^{-5}$  Hartree/Bohr. 
The same elastic boundary conditions as discussed above were used, but here the reference lattice parameter $a_0$ for $\eta=0$ was set to the cubic phase at 0~K.}

\section{Results}
\subsection{BTO}
\begin{figure}
    \centering
    \includegraphics[width=0.48\textwidth, trim=5.3cm 5.6cm 5.8cm 0.8cm, clip]{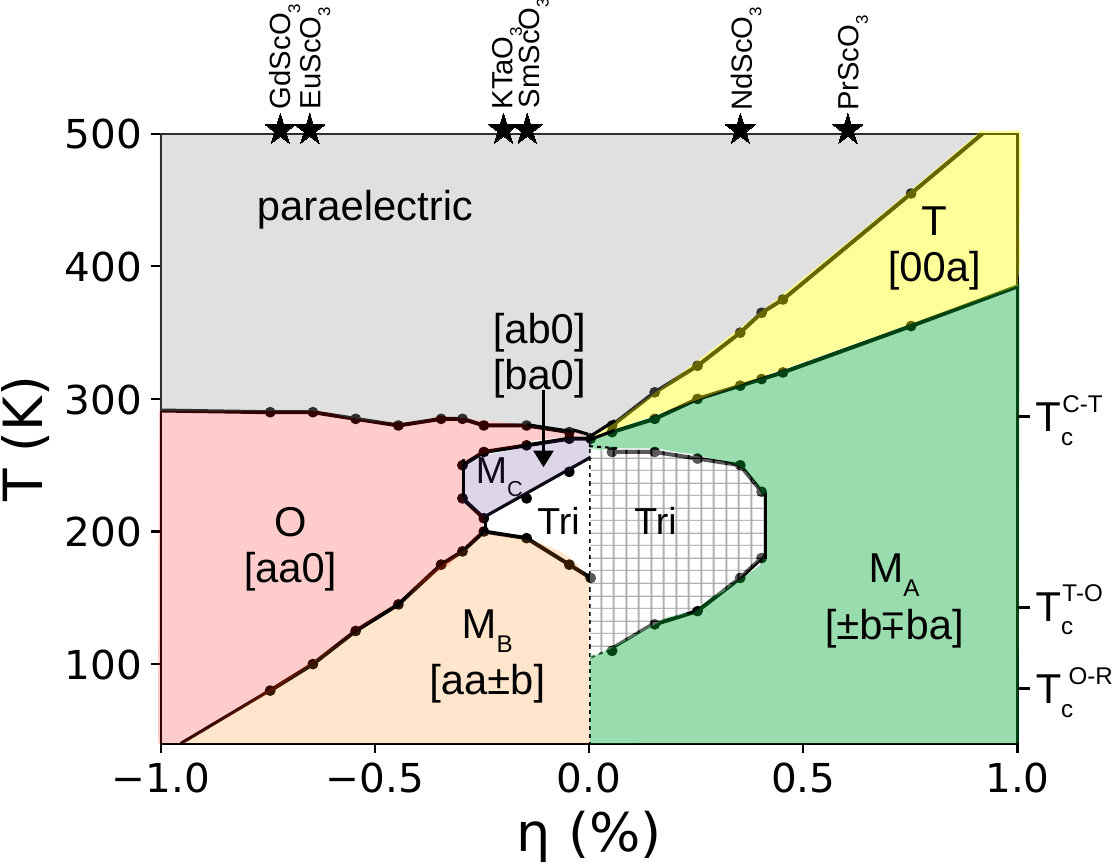}
    \caption{Phase diagram of \bto\ under epitaxial (110) strain.  The bulk \change{transition temperatures} as predicted by our model of 280~K, 140~K, and 80~K for C to T ($T_{c}^{C-T}$), T to O ($T_{c}^{T-O}$), and O to R phase ($T_{c}^{O-R}$)
    are given as reference (right). Observed phases and domain structures are color-coded as: paraelectric (gray), T (${[00a]}$, yellow), M$_\text{A}$ (${[\pm b\mp ba]}$,  green), O (${[aa0]}$, red), M$_\text{C}$ (${[ab0]}$ or ${[ba0]}$, purple), M$_\text{B}$ (${[aa\pm b]}$,  orange), and triclinic (Tri) phases with two different domain structures (white with and without hatches, see text). \change{Black stars at the top of the diagram show the pseudocubic lattice constant for common substrates at room temperature.\cite{schlom_strain_2007}}} 
    \label{fig:bto_phase}
\end{figure}
\begin{figure*}
    \centering
    \begin{overpic}[width=0.85\textwidth, trim=0cm 8cm 3.5cm 0cm,  clip]{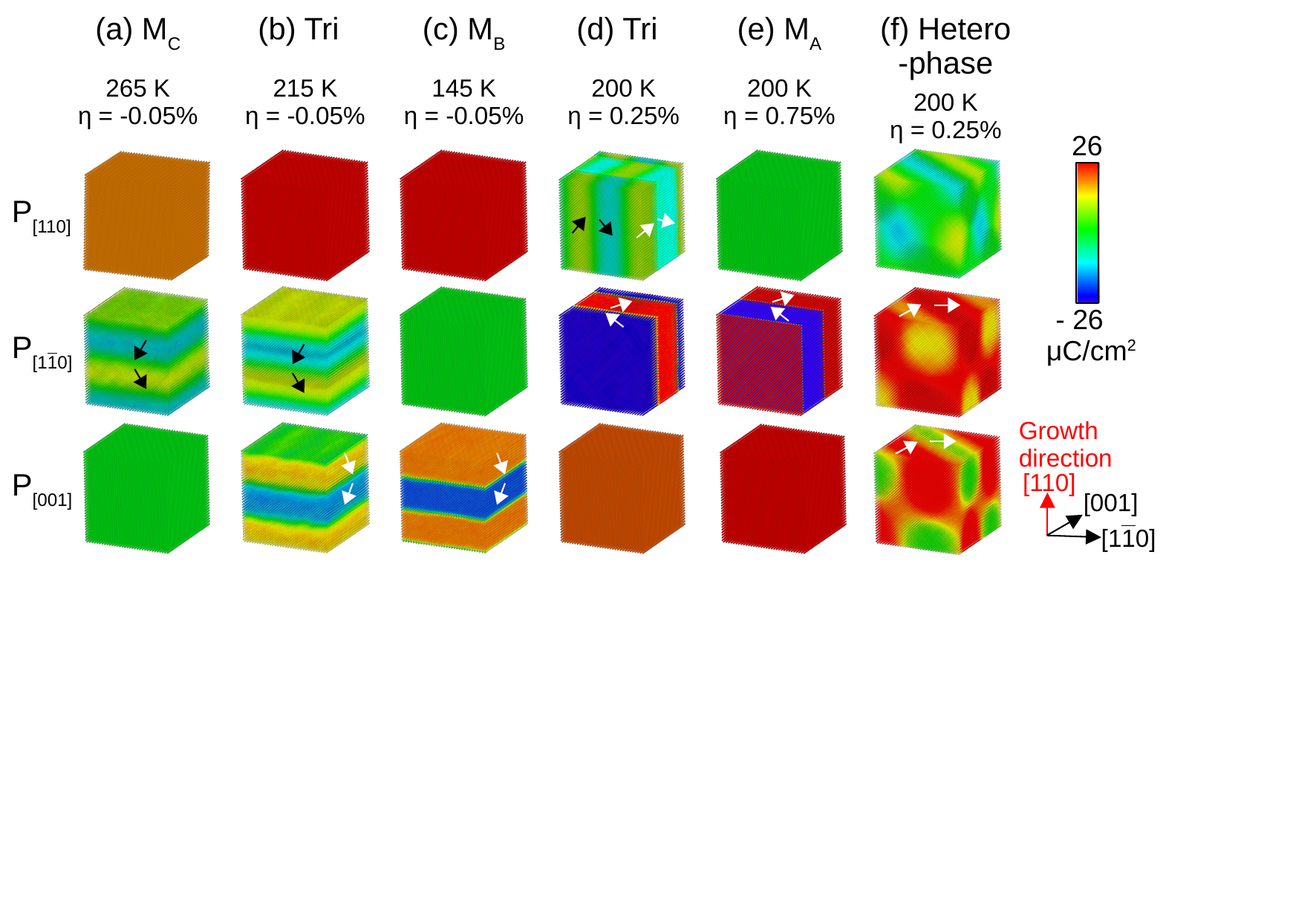}
    \end{overpic}   
    \caption{Collection of metastable domain structures found under \pla{110} strain:  (a)--(e) Structures observed for \bto\ during cooling, cf.~Fig.~\ref{fig:bto_phase},  (f) heterophase phase found during heating of \bto\  from the single domain \ma\ phase. For each structure, the polarization projected on  \pol{110}, \pol{1\bar10}, and \pol{001} direction is shown (rows, color bar: right). Arrows indicate the  polarization of  domains projected on the corresponding  surface planes. The same configurations are found for \kno\ under different conditions.} 
    \label{fig:bto_multi}
\end{figure*}

The phase diagram of \bto\ under epitaxial \pla{110} strain, as shown in Fig.~\ref{fig:bto_phase}, agrees qualitatively with previous theoretical predictions.\cite{wu_effect_2016, ma_effect_2017,wang_variation_2018,gui_properties_2011} However, depending on the chosen parametrization, the phase field models used so far have predicted both negative and positive strain-dependence  of $T_{\text{c}}^{\text{PE-O}}$.\cite{wang_variation_2018} 

On the one hand, compressive strain suppresses \ip\ polarization (\pol{001} and \pol{1\bar10})  and increases the \oop\ lattice constant,  which also stabilizes  \oop\ polarization (\pol{110}). 
\change{Macroscopic \pol{110} can not be realized by tetragonal \pol{100}- and \pol{010}-domains, as they have  unfavourable \ip\ polarization components.} 
Consequently, the system  transforms directly from \pa\ to the O phase (red region in Fig.~\ref{fig:bto_phase}). 
Under cooling, the macroscopic polarization (of the whole system)  remains along \dir{110}.  For $\eta\leq-0.3\%$, additional local \pol{00\pm 1}  nucleates  and the system transforms to a multidomain \mb\ phase (orange region in Fig.~\ref{fig:bto_phase}). 
Under small compressive strain, the other in-plane polarization  nucleates \change{first}, resulting in a multidomain \mc\ phase (purple region) with polarization between O and T, see Fig.~\ref{fig:phasedirection}.  However, the polarization stays close to \pol{110} and the T phase is never reached. 
This \mc\ phase is only present in a small temperature range, and under further cooling, the system transforms first to a 
 Tri phase with \pol{ac\pm b} and \pol{ca \pm b} domains, and then  the multidomain \mb\ phase. This phase is more favourable at low temperatures, as  \change{its polarization  bridges O and R phases and is thus closer} to the ground state of the unstrained material.

On the other hand, tensile strain favours \ip\ polarization (\pol{001} and \pol{1\bar10}). At high temperatures, only the former polarization direction is present (yellow area), as it corresponds to the high-temperature T phase of the bulk.
Cooling down from this phase, the macroscopic polarization stays along the \dir{001} direction and domains with local polarization along the other in-plane direction nucleate. \change{This results} in a multidomain \ma\ phase (green area) with polarization bridging T and R phases, see Fig.~\ref{fig:phasedirection}.
Only under small strain ($\eta\leq 0.4$\%), \change{an additional} multidomain Tri phase (hatched white area) forms near the bulk $T_\text{c}^{\text{C-T}}$. The additional \pol{110} component of the polarization in this phase is smaller than both in-plane ones and reaches a maximum of around 16~\unitp\ at 200~K close to $\eta=0\%$. 
Below $T_\text{c}^\text{O-R}$ of the bulk, the \oop\ component disappears  and the \ma\ phase re-enters.

\change{
The re-entrance of the \ma\ polarization direction can be understood by the  in-plane strain the systems experience at a given temperature. While $\eta$ is defined relative to the paraelectric phase, the actual strain at lower temperatures as compared to the relaxed T, O, or R phase differs, cf.~Fig.~\ref{fig:reentrance}. For small tensile strain and intermediate temperatures, either $a_{[001]}$ or $a_{[1\bar{1}0]}$ is under compression, favouring the Tri-phase with $P_{[110]}$, while at high and low temperatures both directions are under tensile strain and the \ma\  phase is favourable. The re-entrance of polarization directions  at low temperatures has also been predicted by  phase field models.\cite{wu_effect_2016,wang_variation_2018}
}

With increasing tensile strain, i.e.,\ increasing  \ip\ lattice parameters, the transition temperatures where  \ip\ polarization increases abruptly, namely $T_{\text{c}}^{\text{T-M}_\text{A}}$, $T_{\text{c}}^{\text{O-M}_\text{C}}$, $T_\text{c}^\text{PE-T}$,   $T_{\text{c}}^{\text{M}_\text{C}-\text{Tri}}$, and $T_{\text{c}}^{\text{O-M}_\text{B}}$, increase linearly as well. As shown in Fig.~\ref{fig:bto_phase} and Tab.~\ref{tab:slope}, all transitions where \pol{001} increases abruptly (the former two) share a similar  slope of $T_\text{c}(\eta)$, and so do those where \pol{1\bar10} increases abruptly (the latter three). Furthermore, the transition temperature at the lower Tri-\ma\ phase boundary, where the \oop\ polarization vanishes, increases with tensile strain as well.

Following the same logic, with increasing compressive strain, one would expect that the transition temperature where the \oop\ polarization increases abruptly  or where the \ip\ polarization decreases abruptly, namely at the  Tri-\mb, \ma-Tri, and \pa-O  transitions, also increases. Although this is indeed the case, the slopes are small, particularly the one at the \pa-O transition.
In agreement, the energy difference between  the \pa\ and O phases predicted by the DFT  under the same elastic boundary conditions  is not sensitive to the magnitude of $\eta$. \\

In  \mc, Tri, and \mb\ phases (Fig.~\ref{fig:bto_multi}~(a)-(c)) under small compressive strain or low temperatures, the energy of the system is lowered  by the formation of domains with different directions of the in-plane polarization. The domain walls are thereby parallel to the clamped planes (\lbl\ multidomain structures),  and separate either \pol{\pm1\mp10} (in case of \mc), \pol{00\pm1} (\mb), or both  (Tri) types of domains. 
In contrast to that, under tensile strain the  wall normals are on the interface plane (\sbs\ multidomain structures):  \dir{1$\bar1$0}-walls in \ma\ (Fig.~\ref{fig:bto_multi}~(e)) and superimposed \dir{001}- and \dir{1$\bar1$0}-walls in 
  Tri (Fig.~\ref{fig:bto_multi}~(d)).

\begin{figure}
    \centering
    \begin{overpic}[width=0.48\textwidth, trim=0cm 2.5cm 20cm 12.5cm,  clip]{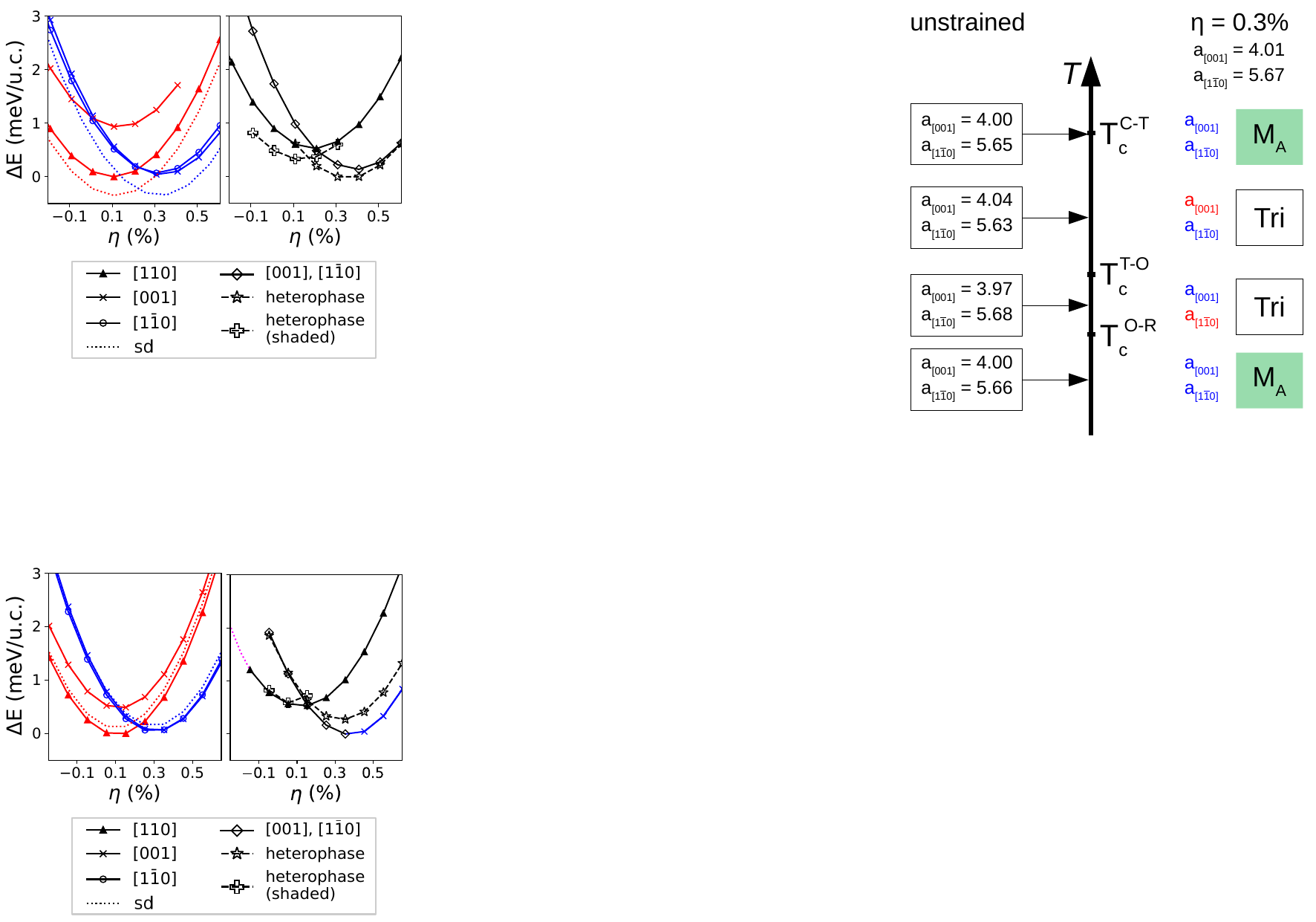}
    \put(26,49){\large \text{(a)}}
    \put(70,49){\large \text{(b)}}
    \end{overpic}
    \caption{Energy as a function of strain $\Delta$E($\eta$) of \bto\ at (a) 45~K and (b) 200~K for possible charge-neutral multidomain configurations (solid curves), single domain states (dotted lines), or heterophases (dashed lines), with \pol{110} and \pol{001} (red), \pol{1\bar10} and \pol{001} (blue), \pol{110}, \pol{001}, and \pol{1\bar10} (black), or  \pol{110} (pink).  Markers indicate the domain wall normal: \lbl\ [110] (triangles),  [1$\bar1$0] (circles), [001] (crosses), as well as [001]- and [1$\bar1$0]-walls (diamonds). Stars and pluses refer to heterophase\change{s with local polarization either close to [1$\bar{1}$0], see Fig.~\ref{fig:bto_multi}~(f) and Fig.~\ref{fig:pdist_hetero}}, or close to [110]. 
    }
    \label{fig:bto_e}
\end{figure}


Although all the phases and domain structures found under cooling can be reproduced  in independent simulations and are (meta)stable during the chosen simulation time, they are  not the only low-energy states at small strain and are not necessarily the energetic ground state of the system at a given strain and temperature. 
Instead, the thermal history, \change{given by} the polarization direction and domain structure \change{at higher temperatures}, matters \change{in cooling simulations}. 
All structures with $\pm$\pol{001} do not form during cooling under tensile strain, as there is a large energy barrier to nucleate the opposite polarization direction in the \sg\ T phase. Analogously, $\pm$\pol{110} domains do not nucleate in the \sg~O phase during cooling under compressive strain.
If the system is cooled down from a phase with \pol{110}, switching to \pol{1\bar{1}0} is also costly in energy. 
This is the reason why there is a vertical phase boundary between these \lbl\ and \sbs\ multidomain structures at $\eta=0$ for all temperatures below the $T_c^{C-T}$, as shown in Fig.~\ref{fig:bto_phase}.  This vertical boundary is not related to an abrupt change of the domain stability, but, instead, caused by the thermal history of the sample.

Thanks to this large energy barrier between the \lbl\ and \sbs\ multidomain structures, \lbl\ ones, which are observed only under compressive strain upon cooling, can be (meta)stable under small tensile strain, and vise versa, see Fig.~\ref{fig:bto_e}~(a). 
Take the \mb\ state \change{at 45~K}  (Fig.~\ref{fig:bto_multi}~(c))  \change{as} an example,  \change{with increasing $\eta$}, the local polarization rotates continuously from \mb\ via R, the ground state in a strain-free material at this low temperature, to \change{another} \ma\ phase \change{with }
\lbl\ \change{domains}, see red triangles in Fig.~\ref{fig:bto_e}~(a).
Similarly, the \ma\ domains observed upon cooling transform (Fig.~\ref{fig:bto_multi}~(e)) remain as \sbs\ structure and transform via R to \mb\ domains, when reducing \change{$\eta$}, see blue crosses in Fig.~\ref{fig:bto_e}~(a). \change{Except for this vertical phase boundary, all the other phase boundaries can be crossed reversibly, e.g., as shown in Fig.~\ref{fig:bto_e}~(b), at 200~K, the polarization of the Tri phases with [110] (black triangles) or with [001] and [1$\bar1$0] (black diamonds) walls rotates continuously to \sg\ O phase (pink dots) or multidomain \ma~(blue crosses) phases.}

There are more low-energy states near the vertical phase boundary, \change{which are charge neutral and (meta)stable once initialized at 45~K but do not form during cooling:}
\sbs\ domains with $\pm$\pol{110} separated by $[001]$-walls (red crosses) and \sbs\ domains with \pol{1\bar10} and [1$\bar1$0]-walls (blue circles).
\change{The latter} even has a slightly lower energy than the \sbs\ state with $\pm$\pol{1\bar10} and \pol{001} (blue crosses), and is the energetic ground state for small  strain up to $\eta \leq 0.22$\%.

Furthermore, as shown in Fig.~\ref{fig:bto_e}~(a),
\change{the local energy minimum of all domain structures is not at $\eta=0\%$ as  the spontaneous R phase would actually be under strain at 45~K for these cubic lattice parameters.}
\change{Instead, configurations with  polarization along \pol{110} and \pol{001} (red), and \pol{1\bar10} and \pol{001}  (blue) reach \pol{aaa} ($\bm{P}=[26,26,26]$ and $\bm{P}=[24,24,24]$) at 0.3\% and 0.1\%, respectively.  At this strain,  the $[001]$- and $\langle 110\rangle$-walls correspond to the typical  R109 and R71 walls of the unstrained material,\cite{marton_domain_2010} respectively.}


All discussed domain configurations found under cooling have lower energy than the \sg\ states. While the relaxation of elastic energy by \sbs\ domains is intuitively expected, actually both \sbs\ and \lbl domains reduce the energy, due to homogeneous strain $V^\text{elas,homo}$ and particularly the strain polarization coupling in $V^\text{coup,homo}$, cf.~Eqn.~\eqref{eq:Eeff}. However, despite the lowering of energy  by the formation of domains, the walls still induce an energy penalty and thus the multidomain states are suppressed in small simulation cells. Thereby, only the minimal numbers of walls  required to fulfill the periodic boundary conditions (2 for \dir{001}-walls and 4 for \dir{110}- or \dir{1$\bar1$0}-walls) nucleates spontaneously.\footnote{Note that the transition temperatures depend on the energy differences between the adjacent phases and thus different domain wall energies may induce finite size effects. However, the impact of this effect on the transition temperatures is minor for the chosen system size. Reducing the simulation box to 36$\times$36$\times$36 unit cells results in a decrease and increase in  $T_{\text{c}}^{\text{O-M}_\text{C}}$ and $T_{\text{c}}^{\text{Tri1-M}_\text{B}}$ by less than 20~K for $\eta=-0.15\%$.}

To explore whether additional (meta)stable configurations may exist, we also heated up the \sg\ \ma\ phase with \pol{b\bar{b}a}.  
Analogous to findings for cooling, there is a high energy barrier to nucleate the opposite polarization direction in the existing polarization components\change{.} Thus, during heating from the \sg\ \ma, instead of forming the Tri phase found under cooling, parts of the system remain in the \ma\ phase  also at higher temperatures, while parts transform to a distorted O phase, resulting in a complex heterophase, see  Fig.~\ref{fig:bto_multi}~(f).
Under further heating, the heterophase transforms to the \sg\ T phase. 
Similarly, when heating up from the \sg\ \mb\ phase with under moderate compressive  strain, a heterophase with a remaining \mb\ phase fraction and a coexisting  \ma\ phase  is found, e.g.,\ between 195~K and 235~K for $\eta=-0.05\%$. \change{Under both compressive and tensile strain, the coexisting \mb\ phase is very close to  O phase, e.g.,\ at $\eta$=0.25\% and  55~K above its formation, the distortion of polarization is 6$^{\circ}$ away from the [1$\bar{1}$0] direction, while the polarization of the coexisting \ma\ phase deviates more from the tetragonal phase, as shown in Fig.~\ref{fig:pdist_hetero}.} As shown in Fig.~\ref{fig:bto_e}~(b),  heterophases which consist distorted O phase with polarization  slightly away from the [1$\bar{1}$0] (black stars) and [110] (black crosses) direction are metastable for $-0.05\% \leq \eta \leq 0.65\%$ and  $-0.05\% \leq \eta \leq 0.15\%$, respectively, and both of them are not lower in energy to the corresponding Tri phases observed upon cooling.

In summary, surprisingly, rich metastable structures and even phase mixtures are found in the chemically simple perovskites \bto. 

\begin{figure}
    \centering
    \includegraphics[width=0.48\textwidth, trim=0.4cm 6cm 10.8cm 0cm, clip]{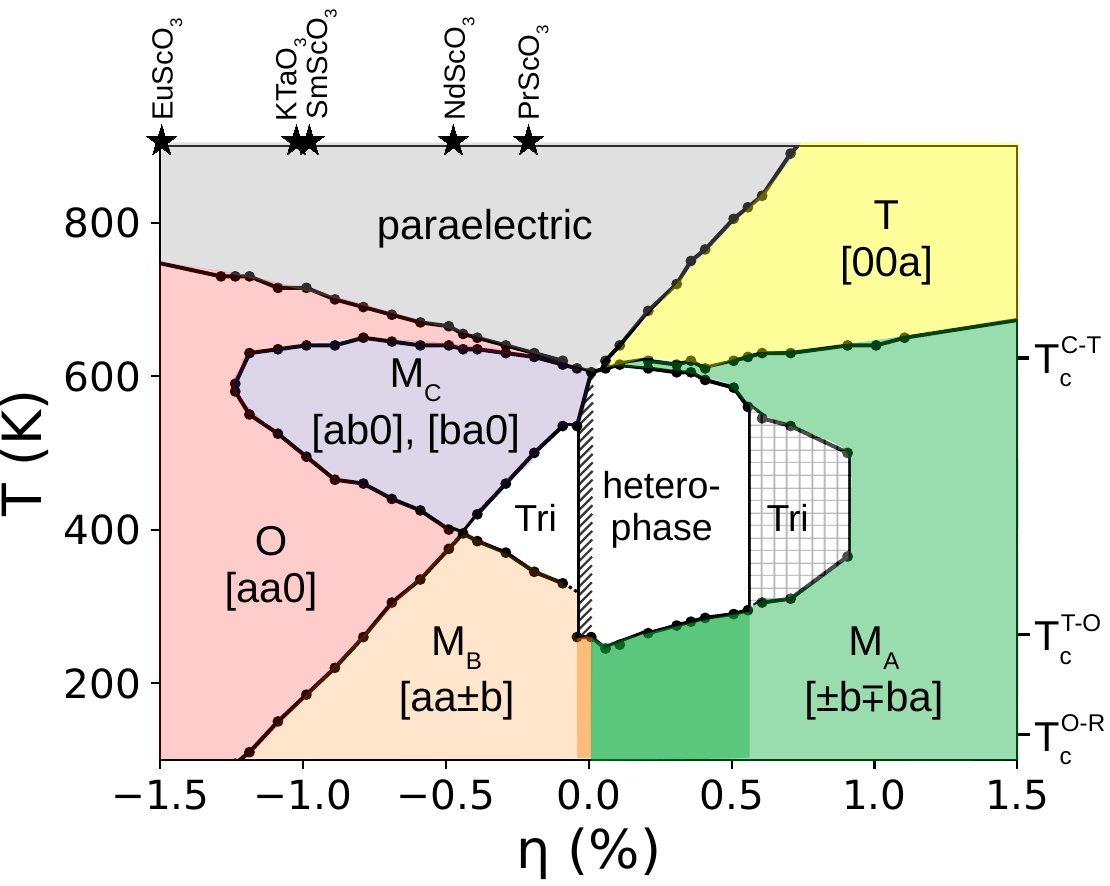}
    \caption{Phase diagram of \kno\ under epitaxial (110) strain. The black dots are the cooling transition temperatures, and are connected through black lines to guide the eye. The bulk $T_{\text{c}}$\change{s} are given as reference (right). Regions of different phases and domain structures are color-coded, with  $\bm{P}$  of the domains annotated: paraelectric (gray), T (${[00a]}$, yellow), M$_\text{A}$ (${[\pm b\mp ba]}$,  green), O (${[aa0]}$, red), M$_\text{C}$ (${[ab0]}$ or ${[ba0]}$, purple), M$_\text{B}$ (${[aa\pm b]}$,  orange), as well as triclinic (Tri) and heterophases each with two different domain \change{configurations}  (see text, white). \change{Black stars at the top of the diagram show the pseudocubic lattice constant for common substrates at room temperature.\cite{schlom_strain_2007}}} 
    \label{fig:kno_phase}
\end{figure}
\subsection{KNO}
\begin{figure}
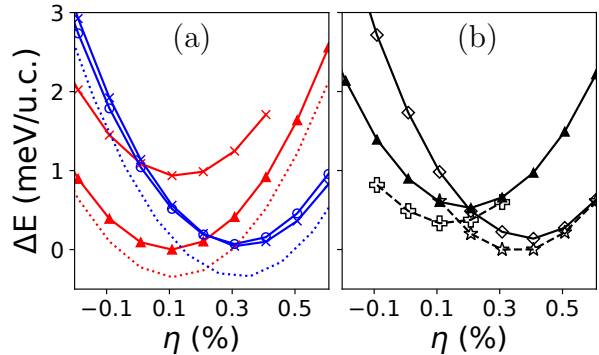

    \centering
    \begin{overpic}[width=0.48\textwidth, trim=0cm 15.2cm 20cm 0cm,  clip]{figures/reentrance_illustration.pdf}
    \put(26,49){\large \text{(a)}}
    \put(70,49){\large \text{(b)}}
    \end{overpic}
    \caption{Energy as a function of strain ($\Delta E$) for \kno\ at (a) 200~K  and (b) 460~K for possible charge-neutral wall configurations (solid curves), single domain states (dotted lines), or heterophases (dashed lines),     with \pol{110} and \pol{001} (red), \pol{1\bar10} and \pol{001} (blue), or with \pol{110}, \pol{001}, and \pol{1\bar10} (black). Markers indicate the wall orientation: [110] (triangles),   [1$\bar1$0] (circles),  [001] (crosses), as well as [001]- and [1$\bar1$0]-walls (diamonds). Stars or pluses refer to heterophase\change{s either with \pol{1\bar10}, see Fig.~\ref{fig:bto_multi}~(f), or \pol{110} regions.} 
    }
    \label{fig:kno_e}
\end{figure}

Unstrained bulk \kno\ shows the same phase sequence as \bto. One may thus expect a similar phase diagram when cooling it down under strain. Indeed, as shown in Fig.~\ref{fig:kno_phase},   O, \mc, \mb, T, \ma, and Tri  
phases with the same domain configurations are observed in the same sequence with respect to temperature and strain. \change{Furthermore,} at $\eta=0$ the \pa-FE transition temperature is equal  to $T_\text{c}^\text{C-T}$  of the bulk, and again  $T_\text{c}^\text{PE-T}$, $T_{\text{c}}^{\text{M}_\text{C}-\text{Tri}}$, and  $T_{\text{c}}^{\text{O-M}_\text{B}}$ also share similar slopes with respect to strain, so do $T_{\text{c}}^{\text{PE-O}}$ and $T_{\text{c}}^{\text{M}_\text{A}-\text{Tri}}$, 
$T_{\text{c}}^{\text{T-M}_\text{A}}$ and $T_{\text{c}}^{\text{O-M}_\text{C}}$,  as well as $T_{\text{c}}^{\text{M}_\text{C}\text{-O}}$ and $T_{\text{c}}^{\text{Tri-M}_\text{B}}$.
Furthermore, as in \bto, the phase diagram also shows an abrupt boundary between configurations with macroscopic \oop\ ($\eta <0$) or \ip\ ($\eta>0$) polarization and the same variety of (meta)stable low-temperature phases is possible close to this boundary, see Fig.~\ref{fig:kno_e}~(a).

Despite these similarities, there are, however, important differences between both materials. First,   \change{the slope} $T_\text{c}(\eta)$ is  almost doubled or halved in \kno\ \change{for the transitions related to abrupt changes of \pol{001} or \pol{1\bar10}}, respectively, and those for \oop\ \pol{110} increase at least by a factor of five, see Tab.~\ref{tab:slope}.
The larger slope of $T_\text{c}^\text{PE-O}$ is also supported by \change{the} energy difference between these two states under the same elastic boundary condition in DFT simulations.
Second, the  \mc\ phase (purple region in Fig.~\ref{fig:kno_phase}) and the phases with all three polarization directions (white regions) occur in a four or two times larger strain range, respectively.

Importantly, in the latter \change{strain} region, the discussed heterophases composed of \mb\ and \ma\ phases form spontaneously during cooling. They are observed between about 600~K and 250~K for $-0.05\% <\eta <0.6\%$, and transform to \sg\ \mb\ or \ma\ phases at lower temperatures under compressive or tensile strain, respectively.
Both, the low-temperature \sg\  phases and the  heterophases,  are  the energetic ground states at 200~K and 460~K, see Fig.~\ref{fig:kno_e}. 

Why are single domain structures and heterophases more favourable in \kno? For both \bto\ and \kno,  the  multidomain phases lower the elastic and elastic coupling energy ($V^\text{elas,homo} + V^\text{coup,homo}$) of the strained films. However, this reduction is smaller for \kno\ and cannot fully compensate the energy penalty for domain wall formation. This can be partly related to the two times smaller  elastic anisotropy in \kno, which can lead to  a smaller energy penalty to distort the single-domain material.\footnote{For the cubic phases of \kno\ and \bto, the anisotropy, $A=2C_{44}/(C_{11}-C_{12})$, calculated based on the elastic constants used in our models are 0.53 and 1.16 which agree with 0.55 (\kno) and 1.28 (\bto) reported in \cite{wang_lattice_2010,wan_structural_2012}.}\cite{ponomareva_influence_2006}

In summary, \kno\ and \bto\ show qualitatively the same trends, but with different transition temperatures and strain ranges of the phases. Besides, related to the two times smaller elastic anisotropy, single domain states and heterophases are more favourable in \kno, while multidomain states of one phase are more favourable in \bto.

\begin{figure}[t]
    \centering
    \includegraphics[width=0.48\textwidth, trim = 0.2cm 5.4cm 9.3cm 0cm, clip]{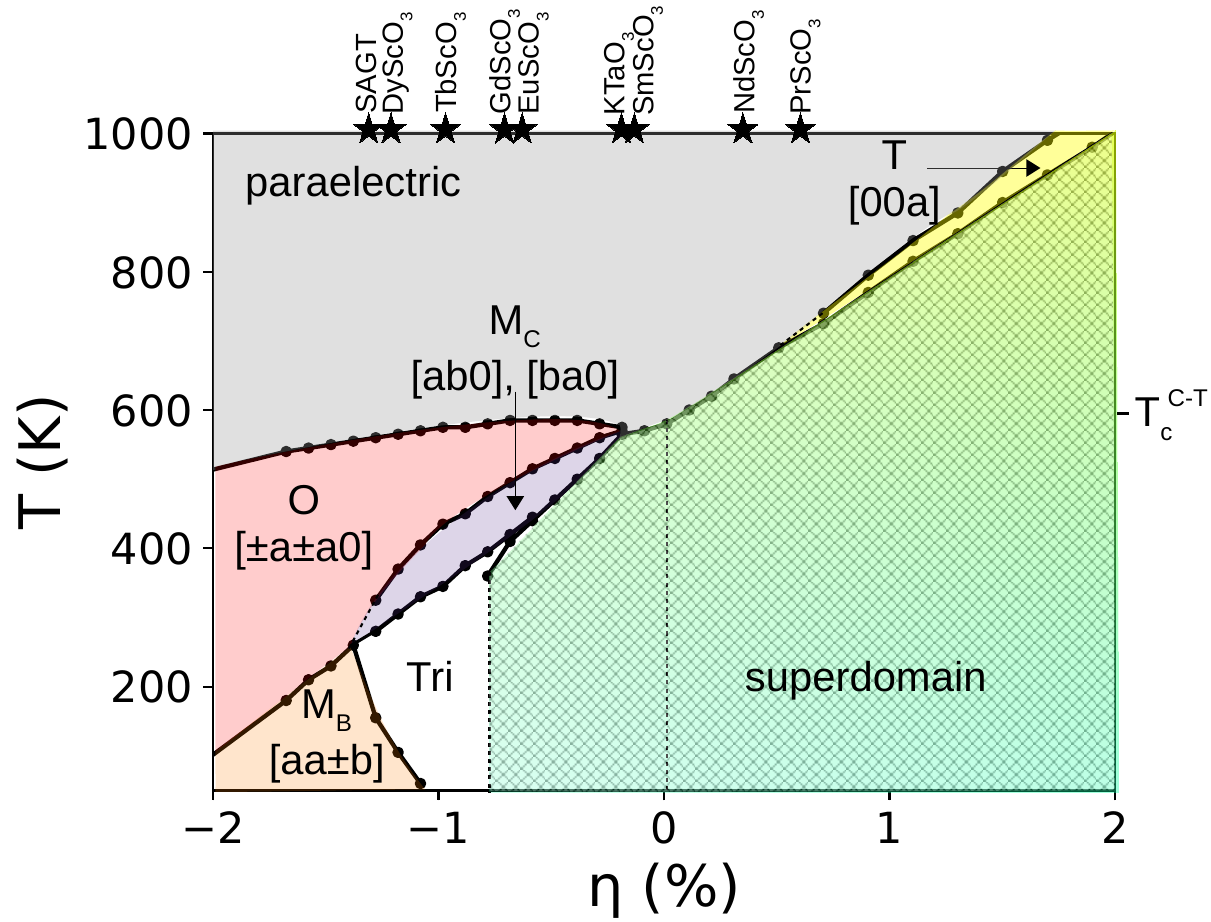}
    \caption{Phase diagram of \pto\ under epitaxial (110) strain. The black dots are the cooling transition temperatures, and are connected through black lines to guide the eye. The bulk $T_{\text{c}}$\change{s} are given as reference (right). Regions of different phases and domain structures are color-coded, with  $\bm{P}$  of the domains annotated: paraelectric (gray), T (${[00a]}$, yellow), O (${[aa0]}$, red), M$_\text{C}$ (${[ab0]}$ or ${[ba0]}$, purple), M$_\text{B}$ ({[aa$\pm{b}$]},  orange),  triclinic (Tri) (see text, white), and superdomain. As indicated by the color gradient, the local polarization in the superdomain state gradually rotates away from T.  \change{Black stars at the top of the diagram show the pseudocubic lattice constant for common substrates at room temperature.\cite{schlom_strain_2007}}} 
    \label{fig:pto_phase}
\end{figure}
\subsection{PTO}

Although the phase diagrams and domain structures of \bto\ and \kno\ are qualitatively similar, subtle differences in their energy landscapes and strain–polarization couplings already lead to noticeable changes. The temperature–strain phase diagram of \pto\ is completely different, see Fig.~\ref{fig:pto_phase}.
The only similarities are that \mc, \mb, and Tri phases and the \sg\ T phase  appear under compressive and  tensile strain, respectively, and that all transitions where \pol{001} changes abruptly share a similar positive slope \change{with strain}, see Tab.~\ref{tab:slope}. 
However, neither heterophases, nor \ma\ phases are observed  under tensile strain, nor is there a vertical phase boundary  at $\eta=0\%$. Furthermore, the T phase with \pol{001} is only stable in a small temperature range under tensile strain (below 30~K at $\eta=1\%$, below our 5~K-resolution for $ \eta < 0.7\%$).

\begin{figure}
    \centering
    \begin{overpic}[width=0.45\textwidth, trim=0cm 14.5cm 18.5cm 1cm, clip]{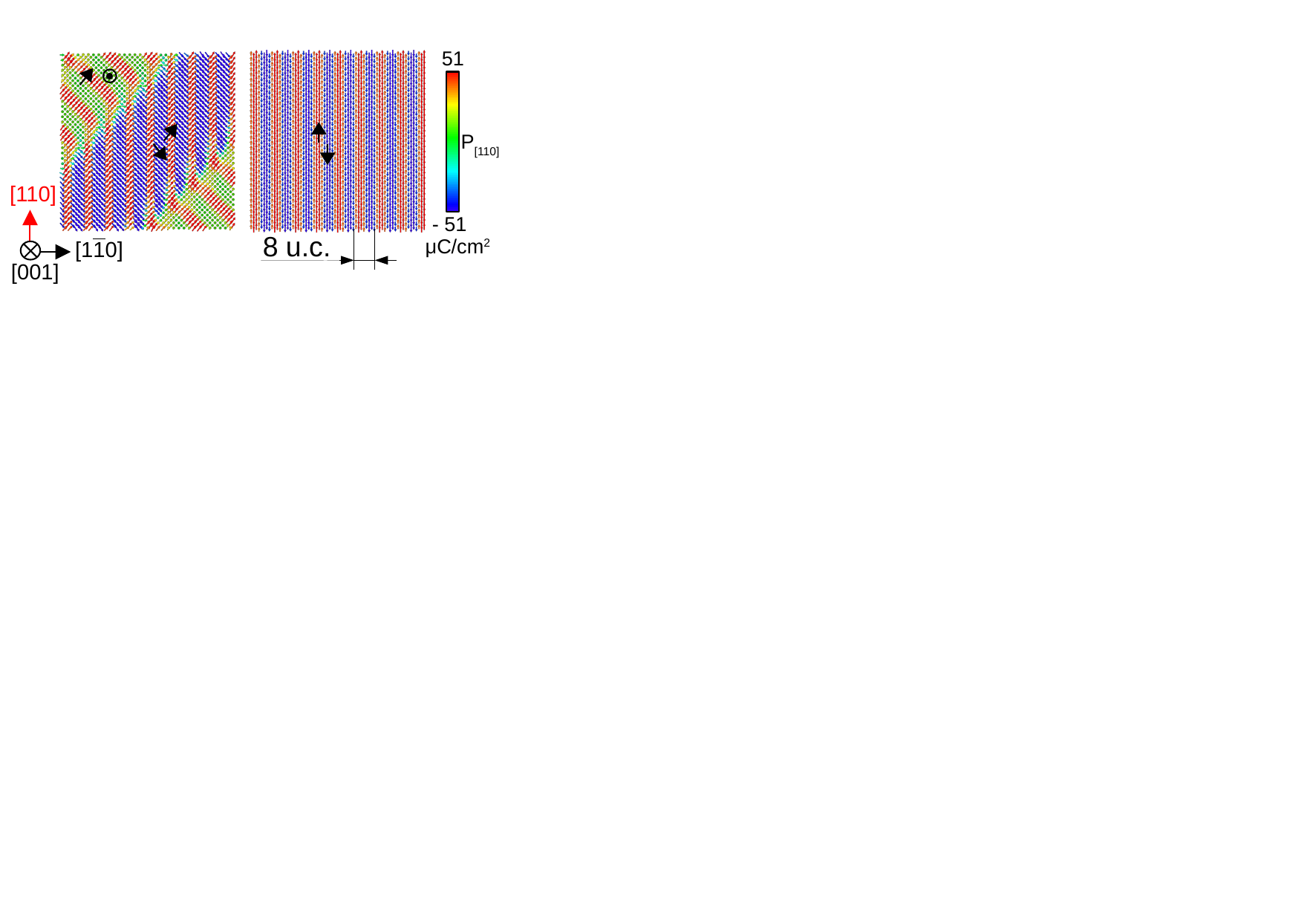}
    \put(12,48){\large \text{(a)}}
    \put(50,48){\large \text{(b)}}
    \end{overpic}
    \caption{Example of the domain structure of \pto\ at 270~K (top view): (a) T-like superdomain state at $\eta=-0.68\%$ and (b) O180\ state with 4 u.c. (unit cell) wide domains at $\eta=-1.48\%$.  Each unit cell is color coded by its \pol{110}. The black arrows on the domain show the polarization directions: $\langle 001 \rangle$ for (a) and $\pm$[110] for (b).}
    \label{fig:pto_multi_top}
\end{figure}

The most important differences to \bto\ and \kno\ are  the complex domain structures. Under tensile strain and down to $\eta=-0.78$\%, the local polarization stays close to $\langle 100 \rangle$ and T90-domains form to accommodate the strain. 
Thereby, a superdomain structure with dense 
 walls  normal to [1$\bar1$0] and normal to [101] form, see Figs.~\ref{fig:pto_multi_top}~(a) and \ref{fig:pdist_superdomain}~(c) for the local polarization configuration and  distribution. Note that approximately the same domain size is found for larger simulations cells (e.g.,\ 12 and 22 [101]-walls \change{for a} system size of $48\times 48\times 48$ and
$96\times 96 \times 96$ u.c., respectively.). At high temperatures, the local polarization \change{direction} is close to the tetragonal \change{axis} (\change{e.g., at 300~K and $\eta=-0.68\%$, the deviation is less than 10$^{\circ}$}).
 These deviations may be partially attributed to the dense domain structures with domain sizes less than 10~u.c. 
Under cooling, \change{the polarization rotation} increases gradually (up to 20$^{\circ}$ at 50~K \change{and $\eta=-0.68\%$}, see Fig.~\ref{fig:pdist_superdomain}) while the domain structure is unchanged. For simplicity, we don't distinguish tetragonal, monoclinic, or triclinic superdomain phases. 
The fact that dense T90 domain \change{walls} occur in \pto\ but not, e.g.,\ in \bto\  is fully in line with large differences in their domain wall energy. \change{DFT predicted that the energy of T90 walls in PbTiO$_3$ is five times lower than the  energy of T180 walls in BaTiO$_3$. The energy of the elastic T90 walls in BaTiO$_3$ is furthermore twice as high as that of the 180 walls.}\cite{grunebohm_domain_2012}

Furthermore, also a large probability to nucleate elastic domain walls at the para--ferroelectric phase   boundary has been reported.\cite{nishimatsu_molecular_2012}

While superdomain structure\change{s have}  already been found in phase field simulations\cite{zhang_strain_2022}, our atomistic simulations reveal for the first time that domains as thin as 16~\AA\ can be stabilized by  (110) strain in \pto.

As T90 domains cannot accommodate compressive (110) strain, \change{compressive} strain induces a phase with local polarization along [110]. However, the \sg\ O phase is not a spontaneous ferroelectric state for bulk \pto, and it is also not even metastable under the strain constraints. 
Instead, an antiferroelectric-like state with local \change{$\pm$}\pol{110} and dense O180-walls along [1$\bar1$0] forms spontaneously, see Fig.~\ref{fig:pto_multi_top}~(\change{b}). \change{The} periodicity of 8~u.c. is an intrinsic property of the material and does not depend on the system size. 
The ultradense \sbs\ O180 walls persist while cooling down, and thus at lower temperatures the  \lbl\ domain structures found in \bto\ and \kno\ do not appear.  Instead, in this antiferroelectric-like state, additional homogeneous polarization components emerge 
 either along \pol{00\pm1} (resulting in \mb) for large compressive strain, or first along \pol{1\bar10} (\mc) and than along both \ip\ directions (Tri). 

\begin{figure}
    \centering
    \includegraphics[width=0.48\textwidth]{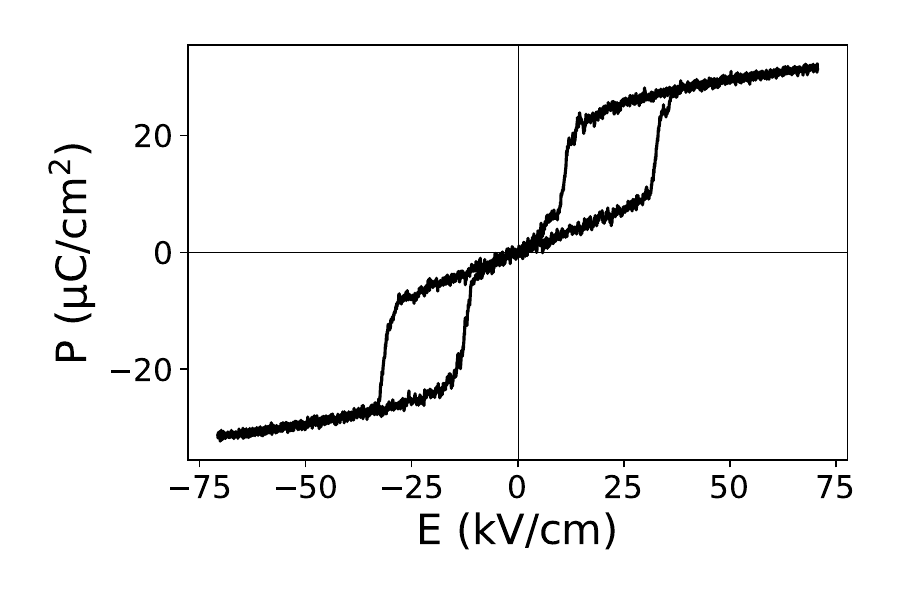}
    \caption{Double loop hysteresis of the antiferroelectric-like O phase for electric field and polarization both along [110] in compressively strained \pto\ ($\eta = -1.47\%$, T = 500~K).}
    \label{fig:pto_afe_doublehysteresis}
\end{figure}

This strain-induced antiferroelectric ordering also shares the functional properties of antiferrolectrics. For example, Fig.~\ref{fig:pto_afe_doublehysteresis}
shows the field response of the state for $\eta = -1.47\%$ and T = 500~K.
For small fields, a linear increase of polarization is induced and at a coercive field of 30~kV/cm the applied field induces the single-domain orthorhombic phase.
This state is not (meta)stable without the applied field and for 15~kV/cm the antiferroelectric state is fully restored. Thus the strain-induced multidomain structure can potentially be used for energy storage applications.

\section{Conclusion}
Based on a  first-principles based effective Hamiltonian, we revealed the impact of epitaxial (110)  strain on 
the strain-temperature phase diagrams of \bto, \kno, and \pto.

Although the phase diagrams and domain structures of \bto\ and \kno\ are qualitatively similar, we showed how the subtle differences in their energy landscapes, \change{ strain–polarization couplings and elastic anisotropy} already lead to noticeable changes in the  phase stability. 
Particularly, \kno\ \change{under strain} favors single domain over multidomain structures.

Most important, the strain along \change{the (110)} low-symmetry direction induces a multitude of complex heterophases and multidomain configurations with nanosized domains,  which cannot be realized by epitaxial (001) strain. \change{ Note that these phases and domain walls are already stabilized by the epitaxial strain in an defect-free material without charged layers or strain gradients. }
Under compressive strain, uncommon \lbl\  domains  with walls parallel to the strained plane form. \change{Especially for small values of strain,  many configurations (with different local polarization and domain structures), including heterophases, are metastable and close in energy in the idealized system. For a macroscopic material with a realistic microstructure, the coexistence of these nearly energy-degenerate configurations, and the large field-induced changes of phase and domain fraction can thus be expected. Notably, large dielectric, piezoelectric, and electrocaloric responses may thus be expected in chemical simple perovskites as the phase boundary is vertical analogous to solid solutions with a morphotropic phase boundary.\cite{roytburd_stability_2011, topolov_heterogeneous_2018, damjanovic_morphotropic_2010, zhou_intermediate_2019}}

In \pto, complex superdomains and nanosize O180 domains were found across a wide temperature-strain range. Both these types of domains may again allow for large functional responses. For example, the
latter corresponds to a strain-induced antiferroelectric-like state and show the typical  double-loop field hysteresis observed in antiferroelectrics. While such antiferroelectric-like response has been reported for thin films in the presence of large depolarization fields \cite{aramberri_ferroelectricparaelectric_2022}, our results show that the strained material already shows this state under short-circuit boundary conditions.

Our findings not only broaden the understanding of strain-mediated ferroelectric behavior, but the diversity of (meta)stable states also suggest design opportunities for adaptive and reconfigurable nanoscale ferroelectric devices by the control of the domain structure by strain, temperature, and thermal history.

\section*{Acknowledgements}
All authors acknowledge financial support from the German research foundation (Deutsche Forschungsgemeinschaft, DFG), Germany (412303109). We thank Chien-Wen Hao for the preliminary testing.

\section*{Data availability}
\change{The  input scripts and post-processed data can be obtained from the authors upon reasonable request. Jupyter notebooks used for analysis can be found in Github\cite{hsu_fighting-mochipublication_epi110_nodate}. }

\bibliography{main}

\section*{Appendix}

\begin{table*}
\caption{Slopes of phase boundaries $T_c(\eta)$ with respect to strain in unit of K/\% for \bto, \kno, and \pto, classified by the abrupt increase (bold) or decrease (non-bold) of the given polarization component. In case of re-entrant transitions, values are given in the second row.}
\begin{tabular}{c|cc|ccc|ccc}
 & \multicolumn{2}{c|}{\pol{110}} & \multicolumn{3}{c|}{\pol{001}} & \multicolumn{3}{c}{\pol{1\bar10}} \\
  & PE--O & \ma--Tri & PE--T & \mc--Tri & O--\mb & T--\ma & O--\mc & Tri--\mb  \\ 
 \hline
\multirow{2}{*}{\bto} & \textbf{-9} & \textbf{-33 }& \textbf{259} & \textbf{200} & \textbf{241} & \textbf{117} & \textbf{67} & -200  \\
 & & 200 &  & &  &  & -301 &  \\ \hline
\multirow{2}{*}{\kno} & \textbf{-95} & \textbf{-150} & \textbf{431} & \textbf{384}  & \textbf{378} & \textbf{54} & \textbf{-17} & -201  \\
 &  & 201 &  & & & & -209 &   \\ \hline
\multirow{2}{*}{\pto} & \textbf{16} &  & \textbf{248} & \textbf{237} & \textbf{252} & & \textbf{230} & -671 
\end{tabular}
\label{tab:slope}
\end{table*}

\begin{figure}
    \centering 
    \begin{overpic}[width=0.35\textwidth, trim=0cm 10.5cm 21cm -1cm, clip]{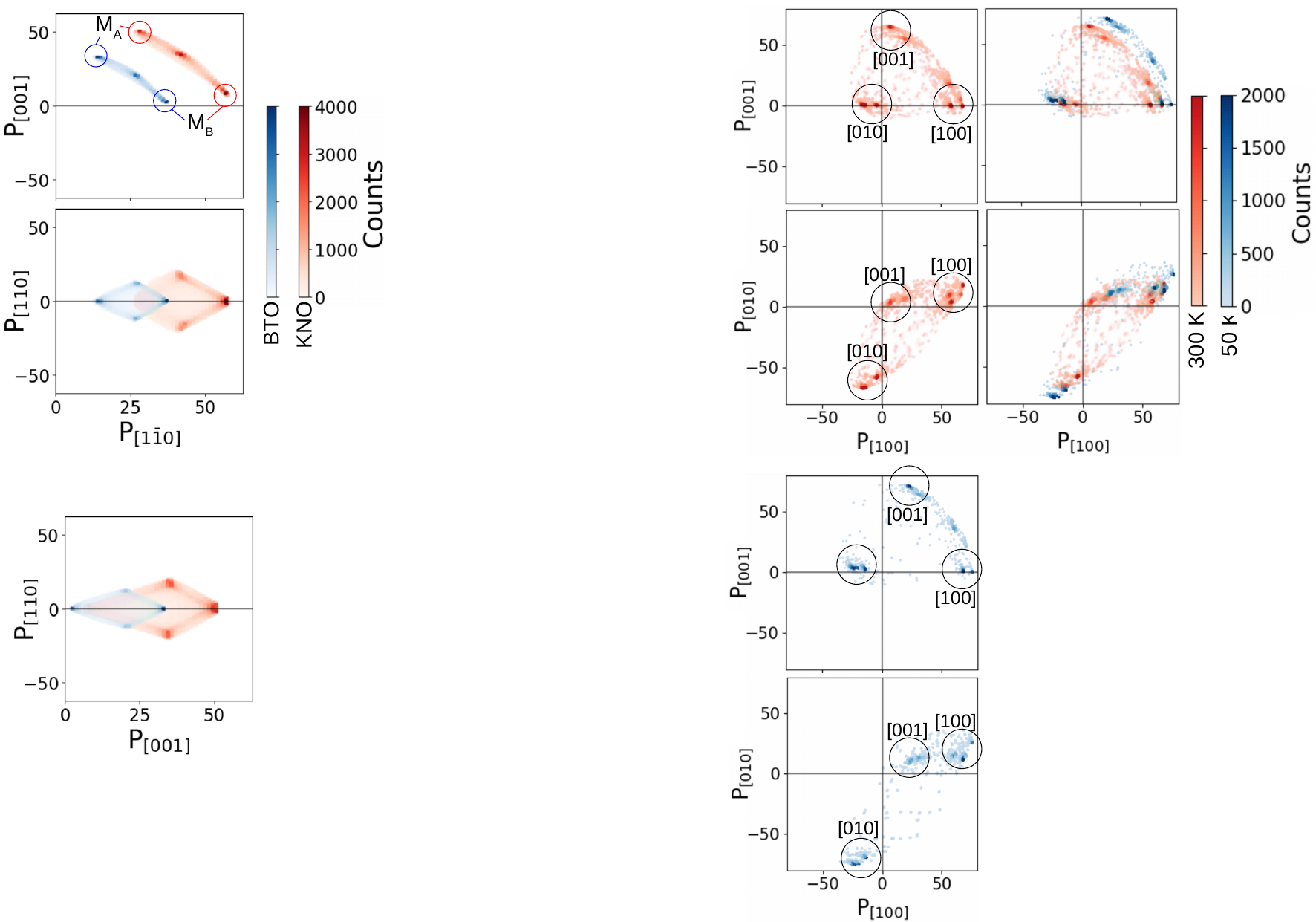}
    \put(13.5,60.5){\text{\large(a)}}
    \put(13.5,18){\text{\large(b)}}
    \end{overpic} 
    \caption{\change{Distribution of polarization} (a) \pol{001} and \pol{1\bar10} as well as (b) \pol{110} in the \ma-\mb\ heterophase configurations observed in \bto\ (blue) and \kno\ (red) under $\eta$=0.25\% at 55~K above their formation upon heating. Circles highlight data from the coexisting \ma\ and \mb\change{, which is only slightly distorted away from O,} domains. The polarization components in unit of $\mu$C/cm$^2$ of each unit cell (dots) are color-coded by their counts.}
    \label{fig:pdist_hetero}
\end{figure}

Figures~\ref{fig:pdist_hetero} shows the distribution of the local polarization in the heterophase for \bto\ and \kno.
\change{Darker colors  correspond to frequent local dipole directions. For \bto\ and \kno\ three clusters of frequent directions can be distinguished in the \pol{1\bar10}-\pol{001} plane, see  Fig.~\ref{fig:pdist_hetero}~(a).  These}
 correspond to the coexisting \ma\ and \mb\ phases (with \pol{001} $>0$  \change{ $\wedge$ \pol{1\bar10} $>$ 0 and  \pol{1\bar10} $>0$ $\wedge$} \pol{001}$\approx 0$), respectively, and the  phase boundaries with finite values of \pol{1\bar10}, \pol{110}, and \pol{001}. \change{In the \pol{1\bar10}-\pol{110} plane four clusters can be distinguished, see}
Fig.~\ref{fig:pdist_hetero}~(b) \change{: Two  with \pol{110}$=0$ for the phases and two for the different phase boundaries shown in the top panel of Fig.~\ref{fig:bto_multi}~(f) (light blue and yellow).} 
\change{Comparing between \bto\ (blue) and \kno\ (red), the polarization of the monoclinic phases in \bto\ stay closer to the corresponding T and O phases, agreeing with \kno's smaller anisotropy: the polarization directions of  \ma\ and \mb\ phases are 9$^{\circ}$ away from [1$\bar1$0] and 30$^{\circ}$ away from [001], while they are 6$^{\circ}$ and 26$^{\circ}$ in \bto, respectively.}

\begin{figure}
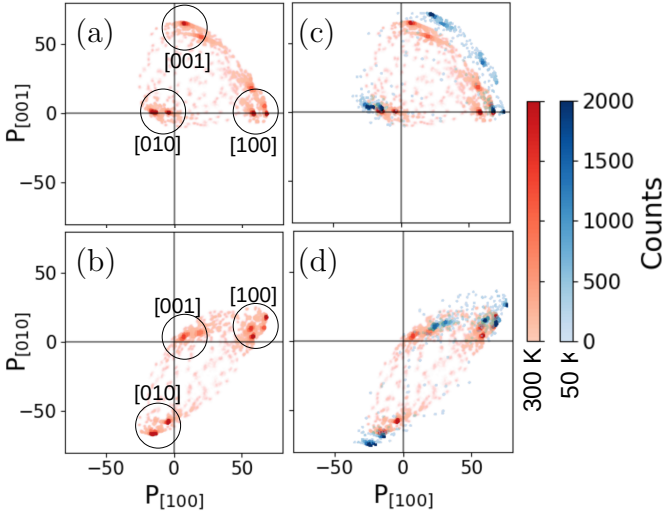

    \centering 
    \begin{overpic}[width=0.5\textwidth, trim=16.5cm 10.5cm 0cm -1cm, clip]{figures/pdist.pdf}
    \put(12,70.5){\text{\large(a)}}
    \put(12,36.5){\text{\large(b)}}
    \put(45,70.5){\text{\large(c)}}
    \put(45,36.5){\text{\large(d)}}
    \end{overpic} 
    \caption{\change{Distribution of polarization}  (a) \pol{001} and \pol{100} as well as (b) \pol{010} and \pol{100} in the superdomain configurations observed in \pto\ at 300~K (red) under $\eta$=-0.68\% upon cooling. Circles highlight data from $\langle 001 \rangle$-like domains. In subfigure (c) and (d),   data at 50~K (blue) is shown as well. The polarization components in unit of $\mu$C/cm$^2$ of each unit cell (dots) are color-coded by their counts. 
    }
    \label{fig:pdist_superdomain}
\end{figure}

\change{Figures~\ref{fig:pdist_hetero}--(a)--(b) show the distribution of the local polarization in the superdomain configuration of \pto at 300~K.}
\change{In both \pol{100}-\pol{001} and \pol{100}-\pol{010} planes, three  local polarization clusters are circled. These clusters have polarization close to the tetragonal directions [100], [0$\bar{1}$0], and [001] and refer to the polarization within the domains. The data points connecting the clusters refer to domain walls.}
As shown in Fig.~\ref{fig:pdist_superdomain}~(c) and (d), \change{the magnitude of the polarization and the distortion of the tetragonal domains increase with decreasing temperature. At 50~K, the angle between P and  [100], [010], and [001] reach about 20$^\circ$, 16$^\circ$, and 21$^\circ$, resulting in domains with Tri-character. }\\

\change{Figure~\ref{fig:reentrance} illustrates the change of actual strain imposed by $\eta=0.3$\% with temperature. Both in-plane directions are under tensile strain relative to the high temperature cubic and the low temperature rhombohedral phases of unstrained \bto. This strain condition stabilizes the \ma\ phase. 
For intermediate temperatures, the lattice is under compressive strain either along $[001]$ or along $[1\bar{1}0]$ compared to the tetragonal and orthorhombic phases. Because of that a finite \oop polarization and an overall Tri phase is induced. Therefore, the \ma phase is present at high temperatures and re-enters at low temperatures.
}\\
\begin{figure}
    \centering
    \includegraphics[width=0.4\textwidth, trim=20.5cm 11cm 0cm 0cm, clip]{figures/reentrance_illustration.pdf}
    \caption{\change{Change of actual strain with temperature underlying the re-entrant 
    transition of the \ma\ phase under small tensile strain in \bto. At $\eta=0.3\%$,  the lattice constant along the [001] and [1$\bar1$0] directions are fixed to 4.01 and 5.67 \AA. Color of the lattice constants tell the real strain the system experience, based on the lattice parameters of the unstrained case at each corresponding temperature: blue or red for tensile or compressive.}}
    \label{fig:reentrance}
\end{figure}

\begin{figure}
    \centering
    \begin{overpic}[width=0.45\textwidth, trim=0.3cm 0cm 20cm 0cm, clip]{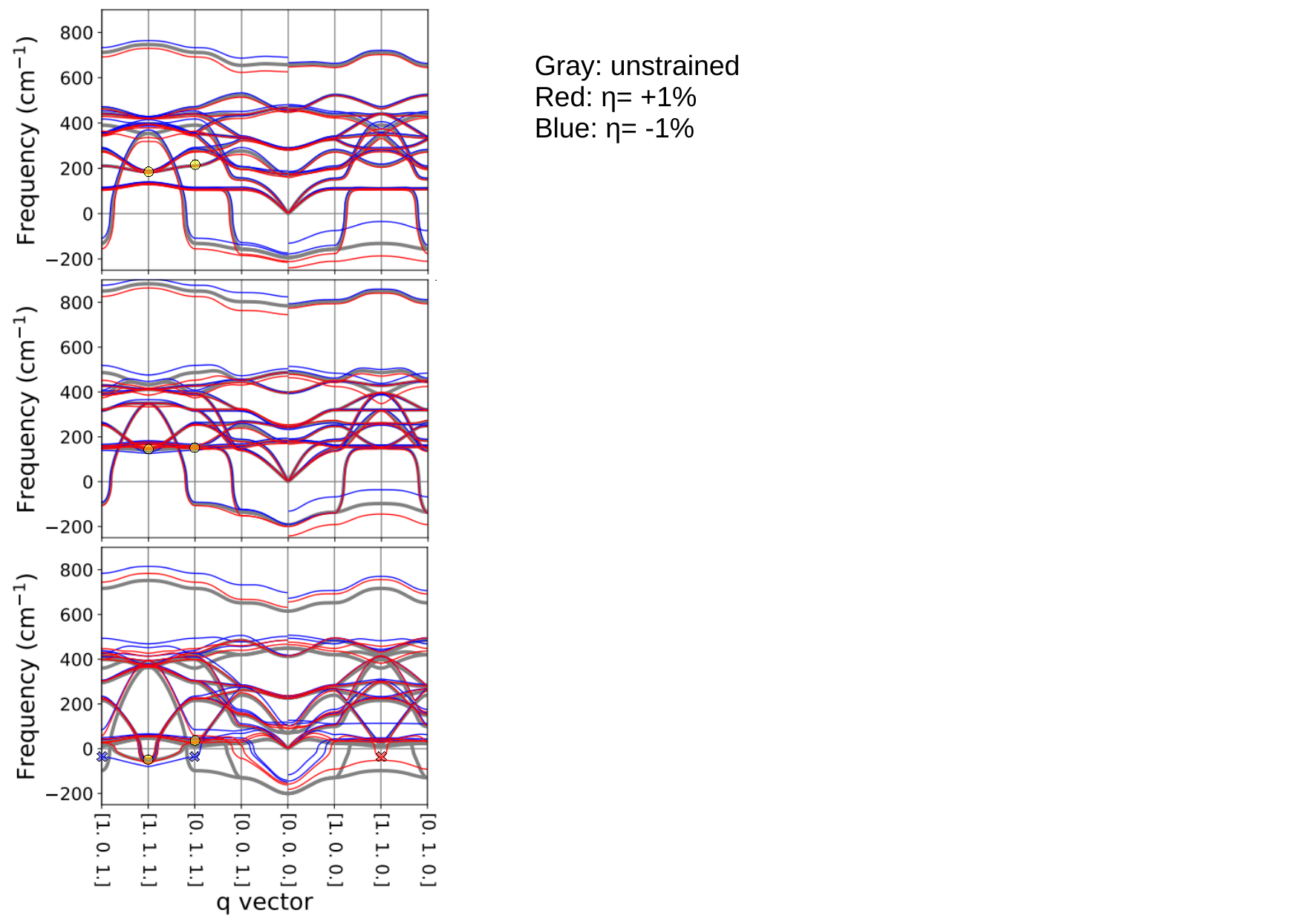}
    \put(11,90){\text{\large(a) BaTiO$_3$}}
    \put(11,63){\text{\large(b) KNbO$_3$}}
    \put(11,33){\text{\large(c) PbTiO$_3$}}
    \end{overpic}
    \caption{The phonon spectra of (a) BaTiO$_3$, (b) KNbO$_3$, and (c) PbTiO$_3$ under tensile $\eta=+1\%$ (red), compressive $\eta=-1\%$ (blue), and zero (thick gray) (110) strain. Yellow circles mark the octahedral rotation mode at $q=\langle 0,1,1 \rangle$ and $\langle 1,1,1 \rangle$ for the unstrained materials.}
    \label{fig:phonon spectrum}
\end{figure}

\change{Figure~\ref{fig:phonon spectrum} compares the phonon spectra of BaTiO$_3$, KNbO$_3$, and PbTiO$_3$ under tensile and compressive (110) strain to that of the unstrained material. 
The phonon spectra of unstrained BaTiO$_3$ and PbTiO$_3$ agree \change{ semi-quantitatively} with literature  [\onlinecite{ghosez_lattice_1999}].

For unstrained  BaTiO$_3$, KNbO$_3$, and  PbTiO$_3$, the dominant soft phonon instabilities are  the polar modes  between $q=[011]$,  $q=[001]$, and $q=[000]$ (Fig.~\ref{fig:phonon spectrum}). 
Strain lifts the 3-fold degeneracy of these polar modes. Atomic shifts along the out-of-plane direction, i.e., [110], (in-plane directions, i.e., [1$\bar{1}$0] and  [001],) are more favorable under compressive (tensile) strain. only in \pto, the ferroelectric instability at q=$[000]$ is reduced under both compressive and tensile (110) strain. As marked by yellow dots,  BaTiO$_3$ KNbO$_3$  are  dynamically stable against octahedral rotations (about  200 cm$^{-1}$ with and without strain). Unstrained PbTiO$_3$ is dynamically unstable against  octahedral rotation (q=$[111]$), however this instability is  by a factor of four  weaker than the ferroelectric instability. 
Furthermore, this octahedral instability is not sensitive to tensile strain and only becomes slightly stronger under compressive strain. Besides, \pto\ is also dynamically unstable against octahedral rotation at $[110]$ under (110)-strain, see the blue and red crosses in Fig.~\ref{fig:phonon spectrum}~(c).  Nevertheless, for the strain range of interest, the ferroelectric instability always dominant.
}

\end{document}